\documentclass[preprpint,5p]{elsarticle}

\usepackage{lineno,hyperref}
\modulolinenumbers[5]

\usepackage{units}
\usepackage{amsmath,amssymb,amsfonts}
\usepackage{xspace}
\usepackage[normalem]{ulem}
\usepackage{xcolor}

\def\Fig#1{Fig.~\ref{#1}}

\def\p{\mathbf{p}}

\def\x{\mathbf{x}}
\def\xp{\mathbf{x}_\perp}

\def\phix{\phi_{\mathbf{x}}}

\newcommand{\eVnospace}{\text{e\kern-0.15ex V}\xspace}
\newcommand{\eV}{\text{ e\kern-0.15ex V}\xspace}

\newcommand{\GeV}{\text{ G\eVnospace}\xspace}

\usepackage{soul}

\journal{Physics Letters B}

\begin{document}

\begin{frontmatter}

\title{Small systems and the single-hit approximation in the \\AMY parton cascade \textsc{Alpaca}}

\author{Robin Törnkvist\fnref{ead1}}
\author{Korinna Zapp\fnref{ead2}}
\fntext[ead1]{robin.tornkvist@hep.lu.se}
\fntext[ead2]{korinna.zapp@hep.lu.se}
\address{Department of Physics, Lund University, Box 118, SE-221 00 Lund, Sweden}

\begin{abstract}

Understanding how momentum anisotropies arise in small collision systems is important for a quantitative understanding of collectivity in terms of QCD dynamics in small and large collision systems. In this letter we present results for small collision systems from the newly developed parton cascade \textsc{Alpaca}, which faithfully encodes the AMY effective kinetic theory. \textsc{Alpaca} reproduces quantitatively previously know results from a calculation in the single-hit approximation for small values of the coupling. We discuss in detail how such a comparison is to be carried out. Particularly at larger coupling a generic differences between the two approaches becomes apparent, namely that in parton cascades particles interact over a finite distance while in direct integrations of the Boltzmann equation the interactions are local. This leads to quantitative differences in the extracted values for the elliptic flow coefficient. These discrepancies appear in situations where the mean free path is not large compared to the interaction time and the applicability of kinetic theory is thus questionable.
\end{abstract}


\end{frontmatter}


\section{Introduction}

The observation of signs of collectivity in small collision systems  \cite{Nagle:2018nvi,Schlichting:2016sqo,Dusling:2015gta} challenges our understanding of both small and large collision systems. In particular, it raises the question to whether there are sizeable final state interactions in small collision systems and to what extent they lead to a (partial) equilibration of the system. Kinetic theory, which successfully describes the early times dynamics and the approach in to equilibrium of large collision systems~\cite{Kurkela:2015qoa} is a promising tool for addressing these questions.

One of the most prominent signs of collective behaviour is the observation of a flow-like momentum anisotropy, quantified by the flow coefficients $v_n$. We concentrate here on the elliptic flow coefficient $v_2$~\cite{ALICE:2014dwt,ATLAS:2013jmi,CMS:2016fnw,PHENIX:2017xrm}, which in the case of collectivity driven by final state interactions is related to an elliptic deformation of the initial state\footnote{It should be kept in mind that final state effects are not the only explanation of non-vanishing $v_n$'s. Correlations in the initial state, e.g. in a CGC picture, can get imprinted on the final state and lead momentum anisotropies that are in agreement with the experimental data in small collision systems~\cite{Schenke:2016lrs}.}. In large collision systems with many interactions per particle one can on general grounds expect that in a kinetic theory description $v_2$ is produce via hydrodynamic-like elliptic flow. In small collision systems with an ellipticially deformed initial state kinetic theory instead produces a momentum anisotropy via the so-called escape mechanism observed both in analytical calculations~\cite{Kurkela:2018ygx,Ambrus:2021fej} and the parton cascade AMPT~\cite{He:2015hfa}.

In this letter we present first results obtained with the parton cascade \textsc{Alpaca} encoding AMY dynamics for small collision systems and discuss in detail how they can be compared to analytical calculations. 

\section{Effective kinetic theory}

The effective kinetic theory (EKT) of QCD at high temperatures was formulated by P.~Arnold, G.~Moore and L.~Yaffe~\cite{Arnold:2002zm}. When $g(T)\ll 1$ the QCD plasma consists of quarks and gluons as well defined quasi-particles with typical momenta $\mathcal{O}(T)$. They propagate as nearly free and nearly massless particles. The EKT describes the dynamics of these hard modes at leading in order in the coupling $\lambda = g^2 N_c$. It is formulated as a set of Boltzmann equations for the spin- and colour-averaged distribution functions $f_s(\mathbf{x}, \mathbf{p},t)$ for particle species $s$
\begin{equation}
    \left(\partial_t + \frac{\p}{p_0} \cdot \nabla_\mathbf{x} \right) f_s(\mathbf{x},\mathbf{p},t) = - C_s^{2\leftrightarrow 2}[f] - C_s^{``1\leftrightarrow 2"}[f] \,.
\end{equation}
The collision kernel $C_s^{2\leftrightarrow 2}[f]$ describes elastic scattering and $C_s^{``1\leftrightarrow 2"}[f]$ contains quasi-collinear spitting and merging. We follow the strategy of~\cite{Kurkela:2021ctp} and extract the elliptic flow coefficient $v_2$ from the energy flow. In the single-hit approximation it is then enough to consider only elastic scattering, as collinear splitting and merging does not effect the energy flow. For consistency and in order not to complicate the discussion we here also restrict ourselves to elastic scattering. Furthermore, since the calculation in~\cite{Kurkela:2021ctp} was done for a purely gluonic system, we do the same here. The elastic scattering kernel has the form
\begin{align}
    & C^{2\leftrightarrow2}[f]  =  \frac{1}{4|\mathbf{p}|} \sum_{bcd} \int_{\mathbf{k}\mathbf{p'}\mathbf{k'}} \delta^{(4)}(P+K-P'-K') \nonumber\\ 
    &  \times \nu |\mathcal{M}^{gg}_{gg} |^2 (2\pi)^4\big\{ f(\mathbf{p})f(\mathbf{k})[1+ f(\mathbf{p'})][1 + f(\mathbf{k'})] \nonumber \\ 
    & - f(\mathbf{p'})f(\mathbf{k'})[1 + f(\mathbf{p})][1 + f(\mathbf{k})] \big\} \,, 
\end{align}
where $|\mathcal{M}^{gg}_{gg}|^2$ is the matrix element for $gg \to gg$ scattering squared summer over the final states and averaged over the initial states, $\nu=16$ is the degeneracy and capital letters denote four-vectors.

For large momentum transfer the matrix element is the same as in vacuum, while for small momentum transfer it receives large corrections due to colour screening. We follow the strategy of ~\cite{York:2014wja,Kurkela:2018oqw}, where it was argued that the main effects of the leading order HTL resummed calculation can be captured by making the replacement
\begin{equation}
    \frac{1}{t^2} \to \frac{1}{(t - \zeta m^2_s)^2} 
\end{equation}
in the vacuum matrix element, where $m_s$ is the effective mass of the parton of spieces $s$ in the propagator and the prefactor $\zeta = e^{5/3}/4$ is chosen to reproduce to leading order the HTL results for drag and momentum broadening ~\cite{York:2014wja,Ghiglieri:2015ala}. 

For isotropic distribution functions the effective gluon mass is given by
\begin{equation}
    \label{eq:mg2}
    m_g^2 = 2 \lambda \int \frac{\mathrm{d}^3\p}{|\p|(2\pi)^3}f(\mathrm{p},\mathrm{x})\,.
\end{equation}
To be  consistent with previous works~\cite{Kurkela:2015qoa,Kurkela:2021ctp} we use this form even for anisotropic distributions.

\section{ALPACA}

The dynamics summarised in the previous section are implemented in the form of a parton cascade in \textsc{Alpaca} (AMY Lorentz invariant PArton CAscade)~\cite{Kurkela:2022qhn}. \textsc{Alpaca} evolves an ensemble of partons in time according to the AMY collision kernels. The partons are treated as massless (thermal masses are $\mathcal{O}(gT)$, i.e.~parametrically small). There is no use of test particles, instead Lorentz invariance is achieved by ordering scatterings, splittings and mergings in a generalised time $\tau$, which is a Lorentz scalar~\cite{Peter:1994yq,Borchers:2000wf}. Interactions occur at discrete instances in $\tau$ and lead to instantaneous changes in the momenta. Between the interactions the partons move on free trajectories. For a detailed discussion of the implementation we refer the reader to~\cite{Kurkela:2022qhn}. In the present context it is only important that elastic scattering between two gluons $i$ and $j$ occurs when their invariant distance $d_{ij}$ at the closest approach of the pair is smaller than $\sqrt{\sigma/\pi}$, where $\sigma$ is the scattering cross section. The invariant distance is given by
\begin{equation}
 d_{ij}  = \sqrt{- (\Delta x)^2 + \frac{(\Delta x_\mu P^\mu)^2}{P^2} } \,, 
\end{equation}
where $\Delta x = x_i-x_j$ is the relative four-distance between the particles, and $P =
p_i+p_j$ is the total four-momentum of the pair.

\section{Single-hit vs. first-hit}

 Following \cite{Kurkela:2021ctp}, we initialize our system with the the Color Glass Condensate-like initial conditions of \cite{Kurkela:2015qoa, Lappi:2011ju} with an added spatial perturbation,
\begin{equation}
    f_{\mathrm{init}} = \frac{A}{p_\xi}e^{-\frac{2p_\xi^2}{3}}\left[1 + \epsilon \frac{|\x_\perp|^2}{R_0^2}\cos\left(2\phix \right) \right]
\end{equation}
where $A$ is the magnitude of the occupancy of the system, $R_0$ corresponds to the system size, $\xi$ controls the momentum asymmetry in the longitudinal direction and $\epsilon$ controls the magnitude of the spatial perturbation. The rescaled momentum $p_\xi$ is defined as

\begin{equation}
    p_\xi = \frac{\sqrt{p_\perp^2 + \xi^2p_z^2}}{Q(\x_\perp)}
\end{equation}
where the characteristic energy scale $Q(\x_\perp)$ can be related to the average transverse momentum as $\langle p_\perp^2 \rangle |_{t=t_0, \x_\perp} = Q^2(\x_\perp)$. We choose $Q(\x_\perp)$ as a Gaussian with width $\sqrt{2}R_0$,

\begin{equation}
    Q(\x_\perp) = Q_0e^{-\frac{|\x_\perp|^2}{4R_0^2}}.
\end{equation}
The initial number of particles in each event is then fixed by the parameters above through
\begin{equation}
    N_{\mathrm{init}} = \nu_g \int d^3\mathbf{x} \int \frac{d^3\mathbf{p}}{(2\pi)^3} f_{\mathrm{init}}(\mathbf{x},\mathbf{p}).
\end{equation}

We also define the eccentricity w.r.t. background energy density of our system to be

\begin{equation}
    \varepsilon_2 \equiv-
    \frac{\int d^2 \xp |\x_\perp|^2 \cos(2\phix) e(\xp)}{\int d^2\xp |\x_\perp|^2 e(\xp)}=-\epsilon,
\end{equation}
and so it comes with a sign change compared to $\epsilon$. 

\smallskip

The elliptic flow coefficient $v_2$ is extracted in the same way as in the single-hit calculation, namely from the correlation of the energy flow with the known reaction plane. In the parton cascade this is given by
\begin{equation}
    v_2 = \frac{\sum_{i \in \text{gluons}} p_{\perp,i} \cos(2 \phi_{\p,i})}{\sum_{i \in \text{gluons}} p_{\perp,i}} \,.
\end{equation}
Since in our initial condition the long axis of the initial position space deformation is along the $x$-axis the $v_2$ defined in this way is negative. But since $\varepsilon_2$ is also negative the response coefficient $v_2/\varepsilon_2$ remains positive.

\medskip

In~\cite{Kurkela:2021ctp} the elliptic flow produced during the evolution of this initial condition was calculated in the single-hit approximation by computing the first correction (linear in the scattering kernel) to the free streaming solution. Thus, the first question that arises is whether the average number of scatterings per particle in the full solution from \textsc{Alpaca} is significantly larger than one. As can be seen in Table~\ref{tab:alpaca_CGC_values} this is indeed the case for all parameter points we studied. In order to assess whether \textsc{Alpaca} is consistent with the single-hit approximation we thus have to artificially reduce the number of scatterings in \textsc{Alpaca}. There are two ways to do this: one can consider only the first scattering of each particle (``first-hit''), or one can randomly reject scatterings to obtain an average number of one scattering per particle (``single-hit''). 

In the single-hit calculation it was also observed that the $v_2$ builds up relatively late during the evolution. We therefore expect that the ``first-hit'' option underestimates $v_2$. This follows from our ordering method, where partons meeting head on will have their closest approach (in terms of $d_{ij}$) later than partons which would not meet at the same point in space. The ``single-hit'' version has the benefit of representing the entire evolution of the system at the cost of allowing individual particles to experience more than one scattering as long as the average is close to one.

\medskip

\begin{figure}[tbp]
    \centering
    \includegraphics[angle=-90,width=0.8\linewidth]{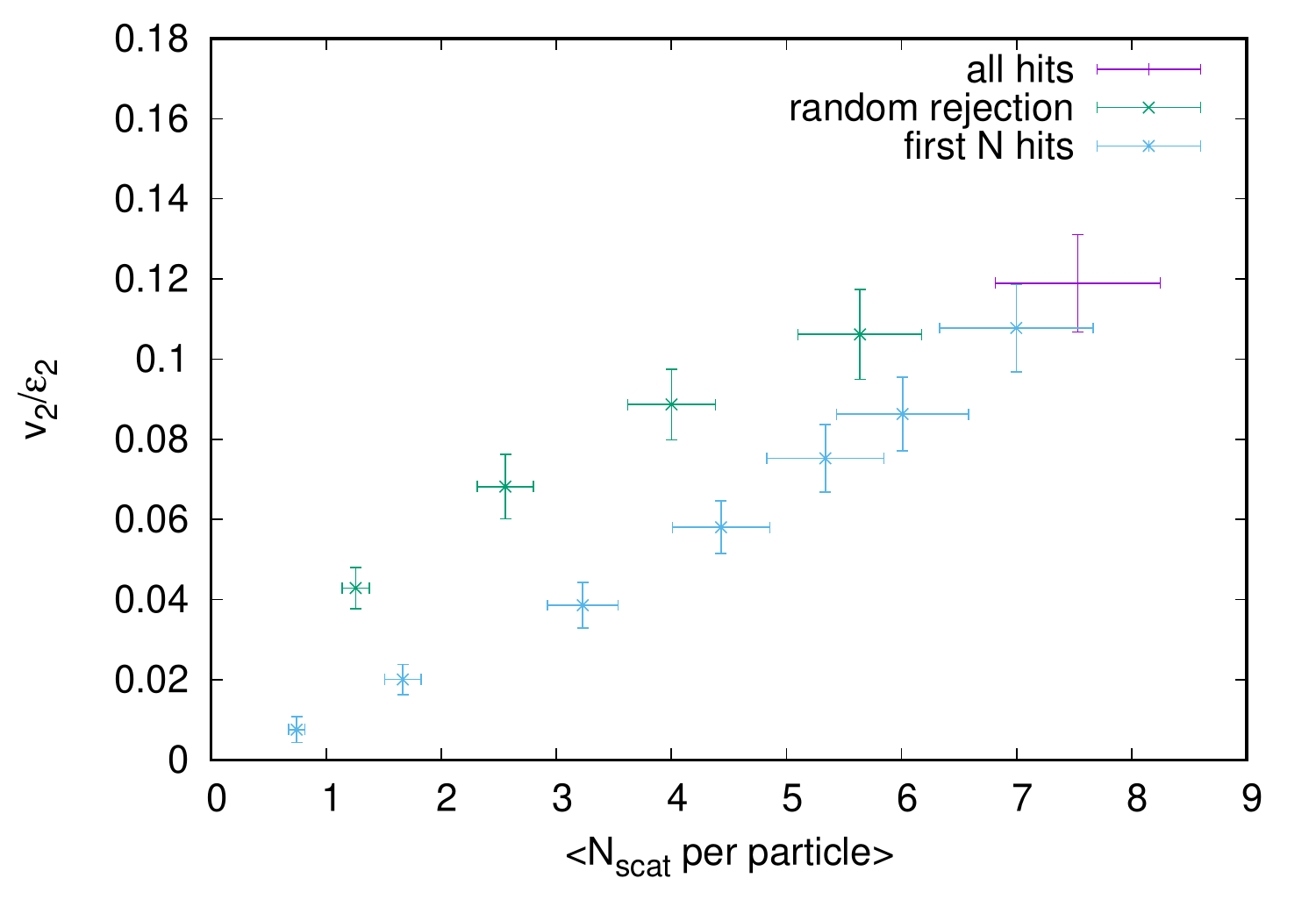}
    \caption{Elliptic flow coefficient $v_2$ vs. numbers of scatterings per particle for first $N$ scatterings and with random rejection of scatterings in the toy model ($\sigma = \unit[5]{GeV^{-2}}$, $A = 0.45$, $Q_0 = \unit[3]{GeV}$, $R_0 = \unit[5]{GeV^{-1}}$, $\xi = 4$, $\epsilon = 0.5$ , $\mu = \unit[1]{GeV}$).}
    \label{fig:v2_vs_Nscat}
\end{figure} 

This discussion does not depend on the AMY dynamics encoded in \textsc{Alpaca}. In order to have a simpler model to demonstrate a few generic effects we work with a toy model. This has the same initial condition and Lorentz invariant evolution as the full calculation but the scattering is simplified. The integrated cross section is set to a constant and the differential cross section is of the form
\begin{equation}
    \frac{\mathrm{d}\,\sigma_\text{toy}}{\mathrm{d}\,t} \propto \frac{1}{(t-\mu^2)^2}
\end{equation}
with a constant regulator $\mu$. Apart from a much shorter run time this toy model has the benefit of decoupling the cross section from the density.

Fig.~\ref{fig:v2_vs_Nscat} shows the generated $v_2$ as a function of number of scatterings for the two ways of reducing the number of scatterings. As expected, the $v_2$ when considering only the first $N$ scatterings is significantly lower than with random rejections of scatterings. A second observation is that $v_2$ first increases with number of scatterings and presumably transitions into a hydro-like regime. This is qualitatively consistent with the behaviour observed in~\cite{Kurkela:2018qeb,Schlichting:2016sqo}. Table~\ref{tab:cgcpoints_toymodel} shows that $v_2$ from first hit is smaller than that from single hit also for other parameters in the toy model.

\begin{figure}[tbp]
    \centering
    \includegraphics[width=0.49\linewidth]{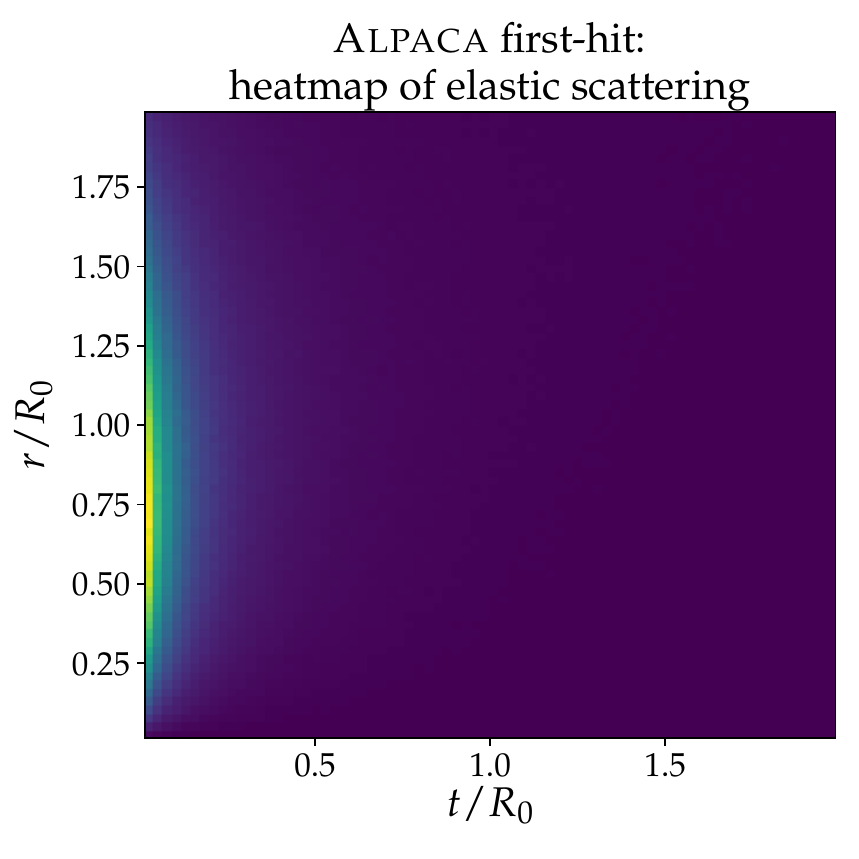}
    \includegraphics[width=0.49\linewidth]{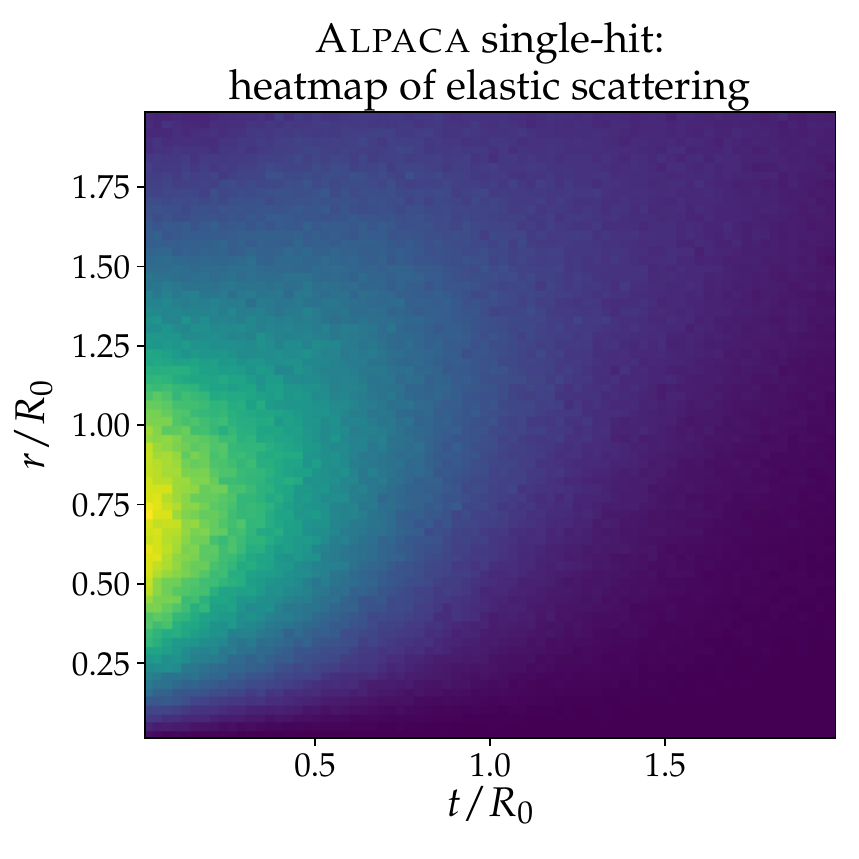}
    \caption{Heatmap of where the elastic scattering in \textsc{Alpaca} takes place in scaled radius $r/R_0$ and time $t/R_0$. The initial parameters are $A=0.45$, $\xi=4$, $Q_0=3$ \GeV, $R_0 = 5$ $\GeV^{-1}$, $\epsilon=0.5$ and $\lambda=5$. A clear difference is observed where the first-hit (left) has most scatterings occurring shortly after initialization, while the single-hit allows for more scatterings at later times while also rejecting some of the scatterings at initialization time.}
    \label{fig:alpaca_elastic_heatmap}
\end{figure} 

Fig.~\ref{fig:alpaca_elastic_heatmap} shows for a calculation in \textsc{Alpaca} where in position and time the scattering takes place when the average number of scatterings is reduced to one. As expected, the scattering happens at very early times in the ``first-hit'' version, leading to an underestimate of $v_2$. The ``single-hit'' option, on the other hand, nicely maps out the entire evolution and thus also captures the expansion of the system. Hence, in the rest of this paper we will focus on the random rejection, which we will refer to as ``single-hit'' in \textsc{Alpaca}.

\begin{table*}
    \begin{tabular}{l|c|c|c|c|c|c|c|c}
    & $A$ & $Q_0$ [$\unit{GeV}$] & $R_0$ [$\unit{GeV^{-1}}$] & $\xi$ & $\mu$  [$\unit{GeV}$]& $v_2/\varepsilon_2$ all hits & $v_2/\varepsilon_2$ first-hit & $v_2/\varepsilon_2$ single-hit \\
    \hline
    \hline
    CGC\,2.3 & 1 & 1.8 & 10 & 4 & 0.894 & 0.19 $\pm$  0.020 & 0.0122  $\pm$  0.0060 &  0.0494 $\pm$  0.0100 \\
    CGC\,2.4 & 0.5 & 1.8 & 10 & 4 & 0.894 & 0.144 $\pm$  0.0160 & 0.0090  $\pm$ 0.0106  & 0.0194  $\pm$  0.0110 \\
    CGC\,3.1 & 1 & 1.8 & 10 & 4 & 1.265 & 0.200  $\pm$  0.022 & 0.0134  $\pm$  0.0056 &  0.0510 $\pm$  0.0092 \\
    CGC\,12.1 & 0.18  & 1.8 & 10 & 4 & 0.187 & 0.046  $\pm$  0.022 & 0.0068 $\pm$  0.0180 & 0.0120  $\pm$  0.0148 \\
    CGC\,12.2 & 0.18  & 1.8 & 10 & 4 & 0.265 & 0.055  $\pm$ 0.015  & 0.0080  $\pm$  0.0182 & 0.0176 $\pm$  0.0166 \\
    CGC\,12.3 & 0.18  & 1.8 & 10 & 4 & 0.374 & 0.066  $\pm$  0.022 & 0.0096  $\pm$  0.0184 & 0.014  $\pm$  0.020 \\
    CGC\,13.1 & 0.18  & 1.8 & 10 & 2.5 & 0.648 &  0.0948 $\pm$ 0.0184 & 0.0100 $\pm$  0.0168 & 0.0266  $\pm$  0.0162 \\
    CGC\,13.2 & 0.18  & 1.8 & 10 & 5 & 0.480 & 0.0632  $\pm$  0.0144 & 0.004  $\pm$  0.026 & 0.008  $\pm$  0.022 \\
    \end{tabular}
    \caption{Comparing $v_2$ values obtained with all hits, first hit and single hit for different parameter points (cf.~\cite{Kurkela:2021ctp}) in the toy model (all points are for $\sigma = \unit[25]{GeV^{-2}}$ and $\epsilon = 0.5$). Note that in the toy model $\mu$ only affects the shape of the differential cross section while the integrated cross section remains fixed. }
    \label{tab:cgcpoints_toymodel}
\end{table*}

\section{Comparing ALPACA to single-hit calculation}

\begin{figure}[tbp]
    \centering
    \includegraphics[width=0.7\linewidth]{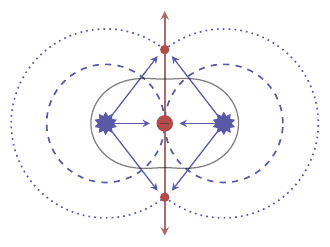}
    \caption{Illustration of the escape mechanism; the production of elliptic flow in the ideal case when scatterings occur locally. We model the initial spatial anisotropy as two hotspots, indicated by blue stars. After some time, the particles streaming horizontally from the hotspots will both have reached the origin, creating an excess of horizontally moving particles there. The AMY kinetic theory locally isotropizes, and so it will remove horizontal movers and create vertical movers, i.e. producing  elliptic flow along the vertical direction. The intersection of particles from the hotspot will move along the vertical axis as time progresses, and since the isotropization occurs in the local rest frame these points will then continuously produce vertical flow.}
    \label{fig:escape_mechanism}
\end{figure} 

\begin{table}[h!]
    \centering
    \begin{tabular}{ l|c|c }
         Param. in \cite{Kurkela:2021ctp} & $\langle N \rangle$  & $\langle d_{ij} \rangle$ $[\mathrm{GeV}^{-1}]$ \\
        \hline		\hline 
        CGC 2.3 & $11.5(2)$ & $21.9(8)$ \\
        \hline
        CGC 2.4 & $15.2(1)$ & $17.0(3)$ \\
        \hline
        CGC 3.1 & $16.8(3)$ & $18.0(6)$ \\
        \hline
        CGC 12.1 & $3.49(5)$ & $10.5(1)$ \\
        \hline
        CGC 12.2 & $5.76(5)$ & $15.6(1)$ \\
        \hline
        CGC 12.3 & $8.47(5)$ & $16.2(2)$ \\
        \hline
        CGC 13.1 & $13.38(7)$ & $15.5(1)$ \\
        \hline
        CGC 13.2 & $10.89(8)$ & $15.3(1)$ \\
  \end{tabular}
      \caption{Resulting data from \textsc{Alpaca} for a subset of the parameter points found in Table I in \cite{Kurkela:2021ctp}, with no restriction in number of scatterings per particle. The parameter $\langle N \rangle$ is the mean number of scatterings per particle and $\langle d_{ij} \rangle$ is the mean invariant distance between the particles that do scatter, at the time of the scattering. All parameter points in the table are initialized with $R_0=10$. The data has been split into ten subsamples, and the errors presented in the parentheses are the errors between the mean of each subsample.}
    \label{tab:alpaca_CGC_values}
\end{table}

To compare the results of elliptic flow produced in \textsc{Alpaca} to those given in \cite{Kurkela:2021ctp}, we restrict ourselves to the parameters for the CGC-like initial conditions which are initialized with  number of partons $N_{\mathrm{initial}} < 500$, for computational reasons.

The $v_2$ presented in \cite{Kurkela:2021ctp} represents the ideal case of local scatterings because in the Boltzmann equation the distribution functions are evaluated at the same position. For small number of interactions per particle $v_2$ is produced via the escape mechanism  explained\footnote{An alternative way of looking at the escape mechanism (e.g.\ in~\cite{He:2015hfa}) is that particles that interact isotropise and thus don't contribute to the final $v_2$. The observed anisotropy is instead generated by the particles that don't interact. It should be noted that the non-interacting particles only have a final state momentum anisotropy when there are particles in the system that do interact and when the interacting particles have an anisotropic momentum distribution before interacting. We would like to point out that this is just a different way of describing the same situation and thus completely equivalent to the explanation we give.} in \Fig{fig:escape_mechanism}. In the simplified situation shown there, the momenta of interacting pairs are completely horizontal in the pairs' centre-of-momentum frame, which maximises the flow signal generated by isotropisation of the momenta of interacting particles. The mechanics of \textsc{Alpaca} do not work in exactly the same way though, as the locality of scatterings depends on an effective cross section $\sigma$ extracted from the AMY collision kernels. Hence, for large values of $\sigma$ the interactions can be very non-local. In \Fig{fig:escape_mechanism} the momenta of interacting pairs are then not exactly horizontal any more in the pairs centre-of-momentum frame and each pair thus creates flow along a different axis. As a result the effective flow created in an event is reduced.

We would like to point out that a sizeable interaction range indicates that we are in a problematic region when it comes to applicability of kinetic theory. The Boltzmann equations rests on the assumption that the interaction time is small compared to the time between scatterings, i.e. the mean free path. Since the typical momentum exchange of a collision is of the order of the screening mass, $m_g$, we expect the interaction time of an elastic collision to parametrically go like $~\sim 1/m_g$. We also know that QCD-like cross sections go like $\sigma \sim 1/m_g^2$, and so the interaction range also scales like $\langle d_{ij} \rangle \sim \sqrt{\sigma} \sim 1/m_g$. When there is approximately one scattering per particle the mean free path is of order of the system size. Thus, when the interaction range is not small compared to the system size it is not small compared to the mean free path, and so neither is the interaction time. We would like to stress, that it is not a priori clear whether in a situation like this the parton cascade approach or the direct integration of the Boltzmann equations is more reliable (or whether both are equally un-reliable).

In Table~\ref{tab:alpaca_CGC_values} we see that \textsc{Alpaca} indeed acts in a non-local way for these parameter points, also with a much larger value of mean scatterings than one. The distribution in $\sigma$ is very wide and contains contributions from many pairs that will never scatter and so the mean is not very telling. Instead we focus on the more relevant quantity $\langle d_{ij} \rangle$, which is the mean distance between the particles that scatter, at the time of scattering. Due to the fact that we find a large $\langle d_{ij} \rangle > R_0$ for all parameter points, we enter a regime where the mean free path is larger than the interaction time, as described above, and so the reliability of the kinetic theory in this range of parameters comes into question. Therefore, we can only quantitatively compare results from \textsc{Alpaca} to the single hit calculation for parameter points where the interaction range is small. Since this is not the case for the parameter points listed in Table~\ref{tab:alpaca_CGC_values} we instead use the scaling formula of \cite{Kurkela:2021ctp}. With this we can make a quantitative comparison between the predicted $v_2$ from the scaling formula in some parameter regime where $d_{ij} < R_0$ (and the initial number of partons is reasonable).

\begin{figure}[tbp]
    \centering
    \includegraphics[width=1\linewidth]{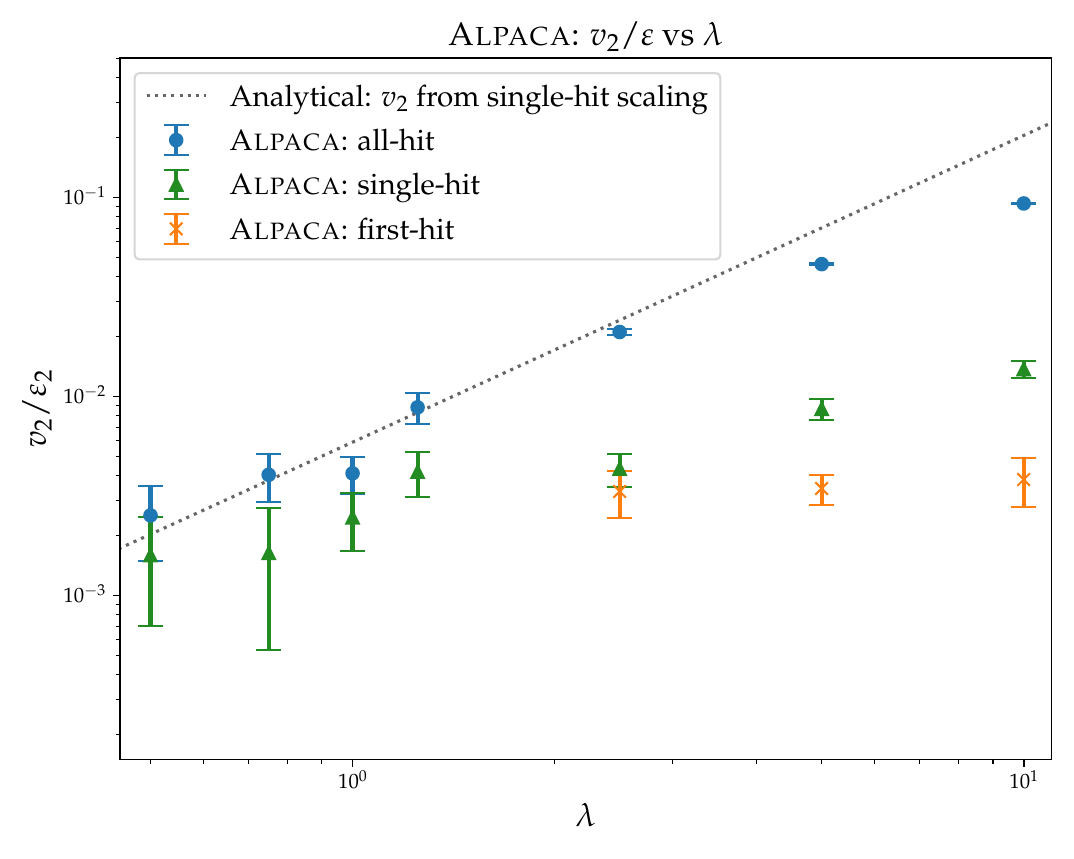}
    \caption{Elliptic flow $v_2$ over the spatial eccentricity $\varepsilon=-\epsilon$, as a function of coupling $\lambda$. The dashed line corresponds to the analytical scaling formula found in \cite{Kurkela:2021ctp}. The marks indicate data from \textsc{Alpaca}. The initial parameters are $A=0.45$, $\xi=4$, $Q_0=3$ \GeV, $R_0 = 5$ $\GeV^{-1}$ and  $\epsilon=0.5$. }
    \label{fig:alpaca_v2_lambda}
\end{figure} 

\begin{table}[h!]
    \centering
    \begin{tabular}{ l|c|c }
         $\lambda$ & $\langle N \rangle$ & $\langle d_{ij} \rangle$ $[\mathrm{GeV}^{-1}]$ \\
        \hline		\hline 
         $0.5$ & $1.66(2)$& $4.22(8)$ \\
        \hline
         $0.75$ & $2.51(3)$ & $6.8(1)$ \\
        \hline
         $1$ & $3.27(4)$ & $9.0(1)$ \\
        \hline
         $1.25$ & $3.99(4)$ & $10.8(1)$ \\
        \hline
         $2.5$ & $6.72(5)$ & $14.5(2)$ \\
        \hline
         $5$ & $10.08(6)$ & $13.09(1)$ \\
        \hline
         $10$ & $13.99(8)$ & $10.31(9)$ \\
  \end{tabular}
      \caption{Resulting data from \textsc{Alpaca} for a subset of the parameter points found in Table I in \cite{Kurkela:2021ctp}, with no restriction in number of scatterings per particle. The parameter $\langle N \rangle$ is the mean number of scatterings per particle and $\langle d_{ij} \rangle$ is the mean invariant distance between the particles that do scatter. All parameter points in the table are initialized with the initial parameters given in \Fig{fig:alpaca_v2_lambda} for varying $\lambda$.}
    \label{tab:alpaca_new_values}
\end{table}

We find such a point at $A=0.45$, $\xi=4$, $Q_0=3$ \GeV, $R_0 = 5$ $\GeV^{-1}$, $\epsilon=0.5$ and $\lambda=0.5$, where $\langle d_{ij}\rangle = 4.22(8)$ $\GeV^{-1}$ and $N_{\mathrm{initial}}=193$. This is still a rather large mean distance of interactions, but it provides a scenario in \textsc{Alpaca} which is closer to the ideal case of completely local scatterings, compared to the parameters shown in Table~\ref{tab:alpaca_CGC_values}. We expand around this parameter in $\lambda$, but note that for $\lambda < 0.5$ a $v_2$ which is distinct from zero proves statistically difficult to find in \textsc{Alpaca}. Hence, we focus on expanding to larger $\lambda$, which will illustrate what happens when we move further away from the localized scattering picture as the cross section increases. In \Fig{fig:alpaca_v2_lambda} the results of this parameter scan is shown, for all-hit (i.e. when no restrictions on number of scatterings is enforced), single-hit and first-hit (with first-hit only being presented for the larger values of $\lambda$ since it proved difficult to extract a value distinct from zero for $\lambda \leq 1.25$). Relevant data corresponding to the all-hit events can be found in Table~\ref{tab:alpaca_new_values}. As can be seen, the starting point of $\lambda=0.5$ agrees reasonably well with the predicted $v_2$ from the scaling formula, both for single-hit and all-hit. The fact that the two results to not differ significantly is expected, since the average number of scatterings for the all-hit scenario is only $\langle N \rangle = 1.66(2)$. As we move to larger $\lambda$ the single-hit results start to diverge from the scaling-formula, due to the scatterings becoming less and less localized. We also observe the same behaviour, but at a slower rate, for the all-hit results. At $\lambda=10$ the mean scattering distance is $\langle d_{ij} \rangle = 10.31$ $\GeV^{-1}$ and so the scatterings are completely non-local, hence the argument of the escape mechanism is not applicable any more. Even with the large amount of scatterings for the all-hit, $\langle N \rangle = 13.99(8)$ we do not reproduce the $v_2$ from the scaling formula. As for the first-hit results, they stay reasonably flat as a function of $\lambda$. This is also expected, since already at $\lambda=2.5$ the mean scattering distance is $\langle d_{ij} \rangle = 14.5$ $\GeV^{-1}$ and so any parton can interact with almost any other parton in the system at the time of initialization, and they will, as can be seen in \Fig{fig:alpaca_elastic_heatmap}. Hence, each parton will expend their one scattering at initial time, in an (on average) non-local way, and then free stream. Since almost no interactions occur at later times when the particles are further away from each other, a larger $\lambda$ and corresponding larger $\sigma$ will show almost no effect.

\section{Conclusions}

Comparing the results of simulations with the AMY parton cascade \textsc{Alpaca} for small collision systems to known results, in particular the single-hit calculation of~\cite{Kurkela:2021ctp} has lead to a number of interesting observations.
\begin{enumerate}
    \item The conclusions in~\cite{Kurkela:2021ctp} that in kinetic theories a sizeable $v_2$ can be obtained from just a single scattering (or less) per particle is confirmed. This has important implications for the interpretation of data from small collision systems.
    \item For the same initial conditions and coupling the number of scatterings per particle is significantly larger than one in \textsc{Alpaca}.
    \item The elliptic flow coefficient $v_2$ increases with the number of scatterings in a way that is qualitatively consistent with the findings in~\cite{Kurkela:2018qeb,Schlichting:2016sqo}.
    \item This raises the question how results from the two calculations can be compared and we show that one has to reduce the number of scatterings in \textsc{Alpaca} by rejecting scatterings with a probability $1/\langle N_\text{scat} \rangle$ to map out the entire evolution of the system.
    \item Generally the elastic scattering cross sections are very large for the parameters considered here. Since the gluons in a parton cascade interact over a finite distance given by $\sqrt{\sigma/\pi}$ the scatterings are sensitive to the gluon distribution in a finite space-time volume. This smears out the $v_2$ obtained via the escape mechanism and leads to a reduction of $v_2$. This implies that a quantitative comparison with the single-hit calculation is only possible for small values of the coupling and our numerical results confirm this. 
    \item The obtained $v_2$ quantifies a clear difference between the approach of solving the Boltzmann equations implicitly through the parton casace \textsc{Alpaca}, and solving them explicitly. These differences appear in a regime where the applicability of kinetic theory is questionable because it is not guaranteed that the mean free path is large compared to the interaction time.
    \item It should also be noted that using test particles in ALPACA would force the interactions to become more local, but this would just bring us back to the case of solving the Boltzmann equations explicitly. Hence, it would not necessarily be more reliable, as we just argued. One should rather regard the difference between the two approaches as a theoretical uncertainty. The fact that \textsc{Alpaca} does not need test particles to avoid causality violation due to frame dependence enables us to quantify this uncertainty for the first time.
\end{enumerate}

We would like to point out that most of these observations are of a general nature and not specific to \textsc{Alpaca} and the single-hit calculation. Instead our considerations apply to other calculations and parton cascades as well.

\section*{Acknowledgments}

We would like to thank Aleksas Mazeliauskas and Aleksi Kurkela for helpful discussions and input.
This study is part of a project that has received funding from the European Research Council (ERC) under the European Union's Horizon 2020 research and innovation programme  (Grant agreement No. 803183, collectiveQCD).

\bibliography{references.bib}

\end{document}